\documentclass[
reprint,
showpacs,
 amsmath,amssymb,
 aps,
]{revtex4-1}

\usepackage{graphicx}
\usepackage{dcolumn}

\begin{document}


\title{Quantum State Transfer in a Two-dimensional Regular Spin Lattice of Triangular Shape}

\author{Hiroshi Miki}
\author{Satoshi Tsujimoto}
\email{miki@amp.i.kyoto-u.ac.jp,tujimoto@i.kyoto-u.ac.jp}
\affiliation{Department of Applied Mathematics and Physics, Graduate School of Informatics, Kyoto University, Sakyo-Ku, Kyoto 606 8501, Japan}

\author{Luc Vinet}
\email{luc.vinet@umontreal.ca}
\affiliation{Centre de recherches math\'{e}matiques, Universit\'{e} de Montr\'{e}al, P. O. Box 6128, Centre-ville Station, Montr\'{e}al (Qu\'{e}bec), H3C 3J7, Canada}

\author{Alexei Zhedanov}
\email{zhedanov@fti.dn.ua}
\affiliation{Donetsk Institute for Physics and Technology, Donetsk 83 114, Ukraine}

\providecommand{\U}[1]{\protect\rule{.1in}{.1in}}
\date{\today}

\begin{abstract}
Quantum state transfer in a triangular domain of a two-dimensional, equally-spaced, spin lattice with non-homogeneous nearest-neighbor couplings is analyzed. An exact solution of the one-excitation dynamics is provided in terms of $2$-variable Krawtchouk orthogonal polynomials that have been recently defined. The probability amplitude for an excitation to transit from one site to another is given. For some values of the parameters, perfect transfer is shown to take place from the apex of the lattice to the boundary hypotenuse.
\end{abstract}


\maketitle

The purpose of this letter is to present an exact solution of the $1$-excitation dynamics of a regular two-dimensional spin-lattice of triangular shape with inhomogeneous nearest-neighbor interactions. It also aims to examine the quantum state transfer properties of this model in the plane. The analytic treatment is proving possible because of recent advances in the theory of multi-variable orthogonal polynomials \cite{Grunbaum,Grunbaum2}. The transfer of quantum states between locations is fundamental in quantum information processing. (See \cite{Bose,Kay} for a review.) Two desirable features are that the transfer be perfect, i.e. realized with probability $1$, and that no external control be required. It is known that spin chains can model such one-dimensional perfect wires with the intrinsic dynamics generating on its own the transfer between the input and output sites with probability $1$ after some time.

The paradigm example \cite{Albanese} is that of the $XX$ chain with Hamiltonian
\begin{equation}\label{1dim-Hamiltonian}
H=\frac{1}{2}\sum_{l=0}^{N-1}J_{l+1} (\sigma_l^x \sigma_{l+1}^x+\sigma_l^y\sigma_{l+1}^y)
\end{equation}
with
\begin{equation}\label{coupling}
J_l=\frac{1}{2}\sqrt{l(N+1-l)}.
\end{equation}
Here $N+1$ is the number of sites and $\sigma_l^x,\sigma_l^y$ (and $\sigma_l^z$) are the Pauli matrices acting at site $l$ on a copy of $\mathbb{C}^2$.
Such $XX$ spin chains with nearest-neighbor couplings preserve the number of excitations and many aspects of quantum state transfer can be studied by restricting to the single-excitation eigenspace.
In this case, the Hamiltonian takes the form of a tri-diagonal (Jacobi) matrix and it can hence be diagonalized by orthogonal polynomials.

The construction of $XX$ spin chains with nearest-neighbor couplings that allow perfect state transfer (PST) is by now well understood \cite{Kay,Vinet} as an inverse spectral problem.
For PST to occur, the $1$-excitation spectrum must be such that successive eigenvalues $x_s,~s=0,1,\cdots, N$ satisfy \cite{Kay}
\begin{equation}\label{spectrum}
x_{s+1}-x_s=\frac{\pi }{T}M_s
\end{equation}   
where $T$ is the time of transfer and where $M_s$ are arbitrary positive odd integers.
Once a spectrum satisfying this condition is given, there is an algorithm \cite{Vinet} to obtain the (non-homogeneous) coupling strengths between sites that determine the $XX$ Hamiltonian. 
The simplest spectrum that satisfies condition (\ref{spectrum}) is the linear one:
\begin{equation}
x_s=s-\frac{N}{2},\quad s=0,1,\cdots ,N.
\end{equation}
The constructive approach of \cite{Vinet} directly shows that the corresponding couplings are those given in (\ref{coupling}) which are recognized to be the recurrence coefficients of the symmetric Krawtchouk polynomials \cite{Koekoek}.
We succeeded in finding the two-dimensional analogue of the Hamiltonian (\ref{1dim-Hamiltonian}). We present it here and study its state transfer properties. 

The letter is organized as follows. We first introduce the two-dimensional lattice Hamiltonian. It is defined on a triangular domain and contains $4$ parameters in addition to the integer $N$ which specifies the number of sites $\binom{N}{2}$. We then provide its one-excitation eigenstates in terms of two-variable Krawtchouk polynomials. This allows to examine transition amplitude for an excitation to move from one state to another. Particular attention is paid to the case where the walk is initiated at the apex of the triangle. For some special values of the parameters, the probability to reach a site of the boundary hypotenuse follows a binomial distribution. Perfect transfer is thus seen to occur between the apex and the set of lattice points on that boundary.

\begin{figure}
\setlength\unitlength{20pt}
\begin{picture}(5,5)(0,0)
\put(0,0){\vector(1,0){4.8}}
\put(0,0){\vector(0,1){4.8}}
\put(0,0){\circle*{.2}}
\put(0,1){\circle*{.2}}
\put(0,2){\circle*{.2}}
\put(0,3){\circle*{.2}}
\put(0,4){\circle*{.2}}
\put(1,0){\circle*{.2}}
\put(1,1){\circle*{.2}}
\put(1,2){\circle*{.2}}
\put(1,3){\circle*{.2}}
\put(2,0){\circle*{.2}}
\put(2,1){\circle*{.2}}
\put(2,2){\circle*{.2}}
\put(3,0){\circle*{.2}}
\put(3,1){\circle*{.2}}
\put(4,0){\circle*{.2}}
\put(0,1){\line(1,-1){.955}}
\put(0,2){\line(1,-1){.955}}
\put(1,1){\line(1,-1){.955}}
\put(0,3){\line(1,-1){.955}}
\put(1,2){\line(1,-1){.955}}
\put(2,1){\line(1,-1){.955}}
\put(0,4){\line(1,-1){.955}}
\put(1,3){\line(1,-1){.955}}
\put(2,2){\line(1,-1){.955}}
\put(3,1){\line(1,-1){.955}}
\end{picture}
\caption{Two-dimensional regular lattice with site labeled by the couples $(i,j),~i,j\in \{ 0,\cdots ,N\}$ satisfying the triangular condition $i+j\le N$.\label{fig1}}
\end{figure}
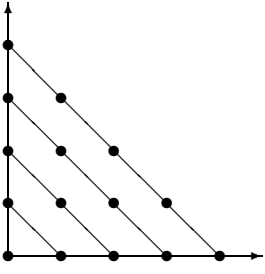

Consider the two-dimensional spin lattice depicted in fig. \ref{fig1}.
Its points are labelled by two indices $(i,j),~i,j\in \{ 0,\cdots ,N\}$ such that $i+j\le N$. The Hamiltonian will be taken of the form
\begin{eqnarray}
H=\sum_{0\le i+j\le N} &&\frac{I_{i+1,j}}{2}(\sigma_{i,j}^x \sigma_{i+1,j}^x+\sigma_{i,j}^y\sigma_{i+1,j}^y) \nonumber \\
+&&\frac{J_{i,j+1}}{2}(\sigma_{i,j}^x \sigma_{i,j+1}^x+\sigma_{i,j}^y\sigma_{i,j+1}^y) \nonumber \\
+&&\frac{1}{2}B_{i,j}(\sigma_{i,j}^z+1)\label{2dimhamiltonian}
\end{eqnarray}
with $I_{0,j}=J_{i,0}=0$ and $I_{i,j}=0=J_{i,j}~i+j>N$.
The constants $I_{i,j}$ and $J_{i,j}$ are couplings the sites $(i-1,j)$ and $(i,j)$ and the sites $(i,j-1)$ and $(i,j)$ respectively. In the last term, $B_{i,j}$ are the strengths of the magnetic fields at the sites $(i,j)$.
Such Hamiltonians are invariant under rotations about the $z$ axis and thus preserve the total number of spins over the lattice that are up (or down):
\begin{equation}
[ H,\sum_{0\le i+j\le N} \sigma_{i,j}^z] = 0.
\end{equation}
We shall now restrict $H$ to the subspace spanned by the states which contain only one excitation, that is only one spin up. A natural basis for that subspace is given by the states
\begin{equation}
\left| i,j\right) = E_{i,j},\quad i,j=0,\cdots ,N
\end{equation}
where $E_{i,j}$ is the $(N+1)\times (N+1)$ dimensional matrix with a $1$ at the $(i,j)$ entry and zeros everywhere else. Note in relation with the action of the Pauli matrices, that the $1$ and the $0$ in this register stand for the vectors
$\binom{1}{0}$ and $\binom{0}{1}$ respectively of the corresponding sites. We shall now look for the one-excitation eigenstates of $H$ denoted by $\left| s,t\right>$ with eigenvalues $x_{s,t}$:
\begin{equation}\label{eigenvalueeq}
H\left| s,t\right> = x_{s,t}\left| s,t\right>.
\end{equation}
Introduce the expansion of $\left| s,t\right>$ in terms of the $\left| i,j\right)$ basis:
\begin{equation}\label{expansion1}
\left| s,t\right> = \sum_{0\le i+j\le N} W_{i,j}(s,t) \left| i,j\right).
\end{equation}
The eigenvalue equation (\ref{eigenvalueeq}) is seen to impose that the coefficients $W_{i,j}(s,t)$ satisfy the $5$-term recurrence relation
\begin{eqnarray}\label{5term}
x_{s,t}W_{i,j}&&=I_{i+1,j}W_{i+1,j}+J_{i,j+1}W_{i,j+1} \nonumber \\
&&+B_{i,j}W_{i,j}+I_{i,j}W_{i-1,j}+J_{i,j}W_{i,j-1}.
\end{eqnarray} 
Let us now fix the coupling constants and the magnetic fields in (\ref{2dimhamiltonian}) to be given by the following expressions which involve the four real parameters $p_1,p_2,p_3$ and $p_4$ and $N$:
\begin{eqnarray}
I_{i,j}&&=\frac{1}{p_1+p_3}\sqrt{ Sp_1p_3(p_2+p_4) i(N+1-i-j)}, \nonumber \\
J_{i,j}&&=-\frac{1}{p_2+p_4}\sqrt{ Sp_2p_4(p_1+p_3) j(N+1-i-j)},\nonumber \\
B_{i,j}&&=\frac{(N-i-j)S}{p_1p_4-p_2p_3}\left\{ \frac{p_2p_4(p_1+p_3)}{p_2+p_4}-\frac{p_1p_3(p_2+p_4)}{p_1+p_3}\right\}  \nonumber \\
&&+j\frac{p_1p_4-p_2p_3}{p_2+p_4}-i\frac{p_1p_4-p_2p_3}{p_1+p_3}
\end{eqnarray} 
with $S=p_1+p_2+p_3+p_4$.
With these coupling constants and magnetic fields, an exact solution is obtained in terms of $2$-variable orthogonal polynomials. Precisely let
\begin{equation}
W_{i,j}(s,t)=W_{0,0}(s,t) \frac{K_{i,j}(s,t)}{\sqrt{r_{i,j}}}
\end{equation}
with 
\begin{equation}
r_{i,j}=\frac{(p_1p_4-p_2p_3)^{2(i+j)}S^{-(i+j)}}{(p_1p_3(p_2+p_4))^i(p_2p_4(p_1+p_3))^j} \binom{N}{i,j}^{-1},
\end{equation}
where $\binom{N}{i,j}=\frac{N!}{i!j!(N-i-j)!}$ is the trinomial distribution function. Eq. (\ref{5term}) now amounts to the following relation for $K_{i,j}(s,t)$
\begin{widetext}
\begin{eqnarray}\label{5termrec}
&&x_{s,t}K_{i,j}(s,t)=(N-i-j)\biggl\{  \frac{p_1p_3(p_2+p_4)S}{(p_1+p_3)(p_1 p_4-p_2p_3)}(K_{i+1,j}(s,t)-K_{i,j}(s,t)) -\frac{p_2p_4(p_1+p_3)S}{(p_2+p_4)(p_1 p_4-p_2 p_3)} \\
&&\cdot (K_{i,j+1}(s,t)-K_{i,j}(s,t)) \biggr\} +i\frac{p_1p_4-p_2p_3}{p_1+p_3}(K_{i-1,j}(s,t)-K_{i,j}(s,t))-j\frac{p_1p_4-p_2p_3}{p_2+p_4}(K_{i,j-1}(s,t)-K_{i,j}(s,t)). \nonumber 
\end{eqnarray}
\end{widetext}
This is recognized as the $5$-term recurrence satisfied by the 2-variable Krawtchouk orthogonal polynomials that have been introduced and characterized recently \cite{Grunbaum,Grunbaum2}. The spectrum is obtained from this identification and thus found to be
\begin{equation}
x_{s,t}=(p_1+p_2)s-(p_3+p_4)t,\quad 0\le s+t\le N,
\end{equation}
a linear expression in $s$ and $t$.
These $2$-variable Krawtchouk polynomials have the following explicit expression:
\begin{widetext}
\begin{equation}
K_{i,j}(s,t)  =\sum _{0\le k+l+m+n\le N}\frac{(-i)_{k+l}(-j)_{m+n}(-s)_{k+m}(-t)_{l+n}}{k!l!m!n!(-N)_{k+l+m+n}}u_1^iv_1^ju_2^kv_2^l
\end{equation}
\end{widetext}
with 
\begin{eqnarray}
u_1&&=\frac{(p_1+p_2)(p_1+p_3)}{p_1S},u_2=\frac{(p_1+p_2)(p_2+p_4)}{p_2S},\nonumber \\
v_1&&=\frac{(p_1+p_3)(p_3+p_4)}{p_3S} ,v_2=\frac{(p_2+p_4)(p_3+p_4)}{p_4S}
\end{eqnarray}
and with $(\alpha )_\beta = \alpha (\alpha+1) \cdots (\alpha +\beta -1) $ the standard Pochhammer symbol.
As eigenstates of a Hermitian Hamiltonian, the vectors $\left| s,t\right>$ can be taken orthonormal:
\begin{equation}
\left< s,t | u,v\right> = \delta_{s,u}\delta_{t,v},
\end{equation}
while the one-excitation vectors manifestly are 
\begin{equation}
\left( i_1,j_1| i_2,j_2\right) = \delta_{i_1,i_2}\delta_{j_1,j_2}.
\end{equation}
This requires the transformation matrix $W_{i,j}(s,t)$ between these basis to be orthogonal. Knowing the weight for which the polynomials $K_{i,j}(s,t)$ are orthogonal, we find that this imposes that
\begin{eqnarray}
&&W_{0,0}(s,t)=\sqrt{\binom{N}{s,t}\frac{1}{[(p_1+p_3)(p_2+p_4)]^N}} \nonumber \\
&&\cdot \sqrt{\frac{(p_1p_4-p_2p_3)^{2(N-s-t)}(p_1p_2)^s(p_3p_4)^t}{(p_1+p_2)^{N-t}(p_3+p_4)^{N-s}}}.
\end{eqnarray}
This completely determines the elements $W_{i,j}(s,t)$.
As a companion to (\ref{expansion1}), in view of the orthogonality of $W_{i,j}(s,t)$, we have the inverse relation
\begin{equation}\label{expansion2}
\left| i,j\right) = \sum_{0\le s+t\le N} W_{i,j}(s,t) \left| s,t\right>.
\end{equation}
We can now write down an explicit expression for the transition amplitude $f_{(i,j),(k,l)}(T)$ for an excitation at the site $(i,j)$ to be found at the site $(k,l)$ after some time $T$:
\begin{eqnarray}\label{transition1}
&&f_{(i,j),(k,l)}(T)=\left( i,j| \exp(-iTH)|k,l\right) \\
&&=\sum_{0\le s+t\le N} W_{i,j}(s,t)W_{k,l}(s,t)e^{iT[(p_3+p_4)t-(p_1+p_2)s]}. \nonumber 
\end{eqnarray} 
(See \cite{Chakrabarti} for analogous calculations in the case of quantum chains.)
From now on we shall restrict the parameters and consider the case where
\begin{equation}\label{restriction1}
p_1=p_4,\quad p_2=p_3.
\end{equation}
With $z\equiv e^{-iT(p_1+p_2)}$, the transition amplitude (\ref{transition1}) simplifies under (\ref{restriction1}) to 
\begin{equation}\label{transition2}
f_{(i,j),(k,l)}(T)=\sum_{0\le s+t\le N} W_{i,j}(s,t)W_{k,l}(s,t)z^{s-t}.
\end{equation} 
These quantities can be calculated with the help of summation formulas \cite{Grunbaum}. In particular, if the quantum walk is initiated at the site $(0,0)$ we have
\begin{eqnarray}\label{apex}
&&f_{(0,0),(i,j)}(T)=\frac{(2p_1p_2(z-1)^2+(p_1+p_2)^2z)^{N-i-j}}{\sqrt{r_{i,j}}} \nonumber \\
&&\cdot \frac{(p_1-p_2)^{i}(z-1)^{i+j}(p_2z+p_1)^i(p_1z+p_2)^j}{(p_2-p_1)^{-j}z^N(p_1+p_2)^{2N}}.
\end{eqnarray}
Now take 
\begin{equation}
T=\frac{\pi}{p_1+p_2}
\end{equation}
(i.e. $z=-1$), the amplitude (\ref{apex}) becomes
\begin{eqnarray}
&&\left| f_{(0,0),(i,j)}\left( \frac{\pi }{p_1+p_2}\right)\right| = \sqrt{ (2p_1p_2)^{i+j}\binom{N}{i,j}} \nonumber \\
&&\cdot \frac{|p_1-p_2|^{(i+j)}}{(p_1+p_2)^{2N}}\left| 8p_1p_2-(p_1+p_2)^2\right| ^{N-i-j}
\end{eqnarray}
Imposing further that 
\begin{equation}\label{restriction2}
(p_1+p_2)^2=8p_1p_2,
\end{equation}
one arrives the following simple expression
\begin{equation}
\left| f_{(0,0),(i,j)}\left( \frac{\pi }{p_1+p_2}\right) \right| =\sqrt{2^{-N}\binom{N}{i}}\delta_{i+j,N}.
\end{equation}
The output excitation will hence distribute binomially on the site of the boundary hypotenuse. It is moreover easy to see therefore that
\begin{equation}
\sum_{k=0}^N \left| f_{(0,0),(k,N-k)}\left( \frac{\pi }{p_1+p_2}\right) \right|^2=1.
\end{equation}  
We thus observe that with these values of the parameters, the Hamiltonian will evolve the state $\left| 0,0\right) $ in time $T=\frac{\pi }{p_1+p_2}$ to any one of the states $\left| i,N-i\right)$ with probability one.
This may be looked at as some kind of perfect transfer. As a consequence we have that
\begin{equation}
\left| f_{(0,0),(i,j)}\left( \frac{\pi }{p_1+p_2}\right) \right|  =0 ,\quad i+j<N.
\end{equation}
This is observed to generalize to 
\begin{equation}
\left| f_{(i,j),(k,l)}\left( \frac{\pi }{p_1+p_2}\right) \right| =0,\quad i+j+k+l<N.
\end{equation}

It is interesting to see what is the form of the Hamiltonian when the choices (\ref{restriction1}) and (\ref{restriction2}) are made for the parameters. One finds that 
the coupling constants $I_{i,j}$ and $J_{i,j}$ and the strengths $B_{i,j}$ of the magnetic field are given by the following simple expression:
\begin{eqnarray}
\frac{I_{i,j}}{p_1+p_2}&&=\frac{1}{2}\sqrt{i(N+1-i-j)}, \nonumber \\
\frac{J_{i,j}}{p_1+p_2}&&=-\frac{1}{2}\sqrt{j(N+1-i-j)},\nonumber \\
\frac{B_{i,j}}{p_1+p_2}&&=\frac{p_1-p_2}{p_1+p_2} (j-i).
\end{eqnarray} 
Here it should be noted that (\ref{restriction2}) implies that
\begin{equation}
\frac{p_1-p_2}{p_1+p_2}=\pm \frac{1}{\sqrt{2}}.
\end{equation}
We see that the magnetic fields are antisymmetric with respect to the lattice diagonal and we observe the symmetry $I_{i,j}\leftrightarrow -J_{i,j}$ under $i\leftrightarrow j$. When this specialization of the parameters is examined at the level of the $2$-variable Krawtchouk polynomials, symmetric functions in $s$ and $t$ result.

Let us summarize to conclude. We have presented a spin lattice which is the two-dimensional analogue of the simplest non-homogeneous quantum spin chain that exhibit perfect state transfer. In parallel to the one-dimensional case, the one-excitation Hamiltonian is diagonalized by two-variable Krawtchouk polynomials. This exact solution allows to show that there is some kind of (generalized) perfect state transfer between one input site and a set of output sites. One can imagine various ways in which this could prove useful. Very few spin lattices are known to be analytically solvable \cite{Kitaev}. Although we only have provided an exact determination of the one-excitation dynamics, we trust that this addition will allow to further explore phenomena inherent to two (higher) dimensional spin networks. 

\begin{acknowledgments}
One of us (H.M.) wishes to thank the CRM for its hospitality
while this work was carried out and gratefully acknowledges a Grant-in-Aid for Japan Society for the Promotion of Science (JSPS)
Fellows. The research of S.T. is funded in part by KAKENHI (22540224). L.V. is supported in part through funds
provided by the National Sciences and Engineering Research Council (NSERC) of Canada.
\end{acknowledgments}

\bibliographystyle{unsrt}
\bibliography{2variable-pst}

\end{document}